\documentclass[runningheads]{llncs}
\usepackage{graphicx}
\usepackage{cite}
\usepackage{url}
\usepackage[bottom]{footmisc}
\usepackage [autostyle]{csquotes}    
\MakeOuterQuote{"}

\usepackage{multirow}
\usepackage{array}
\usepackage{lmodern}
\usepackage{textcomp}
\usepackage[utf8]{inputenc}
\usepackage{longtable}
\usepackage[labelfont=bf]{caption}

\newcommand{\tabitem}{\textbullet~~}
\usepackage{enumitem}
\newcolumntype{C}[1]{>{\centering\arraybackslash}p{#1}}

\begin{document}
\title {An Example of Privacy and Data Protection Best Practices for Biometrics Data Processing in Border Control: Lesson Learned from SMILE} 
\titlerunning{An Example of Privacy and Data Protection Best Practices}
%
\author{Mohamed Abomhara \and Sule Yildirim Yayilgan}
\authorrunning{Mohamed Abomhara et al.}
%
\institute{Department of Information Security and Communication Technology, Norwegian University of Science and Technology, Norway \\
\email{\{mohamed.abomhara, sule.yildirim\}@ntnu.no}}
\maketitle              
\begin{abstract}
Biometric recognition is a highly adopted technology to support different kinds of applications, ranging from security and access control applications to low enforcement applications.  However, such systems raise serious privacy and data protection concerns. Misuse of data, compromising the privacy of individuals and/or authorized processing of data may be irreversible and could have severe consequences on the individual's rights to privacy and data protection. This is partly due to the lack of methods and guidance for the integration of data protection and privacy by design in the system development process. In this paper, we present an example of privacy and data protection best practices to provide more guidance for data controllers and developers on how
to comply with the legal obligation for data protection. These privacy and data protection best practices and considerations are based on the lessons learned from the SMart mobILity at the European land borders (SMILE) project.

\end{abstract}

\section{Introduction}
\label{sec:introduction}

The dramatic advances in computerization and personal data collection have opened doors to unprecedented opportunities in the field of law enforcement and national security applications. However, the increase in data collection, processing, retention and analysis is leading to increased surveillance and tracking of people (data subjects) in many ways \cite{willoughby2017biometric}. When data about individual's activities is collected and analyzed, it can lead to some challenges and conflicts with fundamental human rights and can be the cause of ethical, social and legal challenges such as unauthorized and inadvertent disclosure, embarrassment and harassment, social stigma and inappropriate decisions, to name a few \cite{abomhara2019border,memon2017biometric,sutrop2010ethical,campisi2013security}.  The key challenge and the focus of this paper is related to the respect for individual privacy and the right to personal data protection.

The amount of personal data processed in the border control and its research context continues to rise every year. SMILE\footnote{https://smile-h2020.eu/smile/} as a proposed border control tool can help to optimize and monitor the flows of people at land borders to increase security and improve border crossing efficiency as well as to facilitate effective migration control and enforcement. Like others, SMILE also increases the tendency to collect, use and process sensitive personal data (e.g., alpha-numeric data and biometric data). Thus, SMILE system raises serious legal and policy concerns \cite{abomhara2019border,abomhara2019}. Misuse of data, compromising the privacy of individuals and/or authorized processing of data may be irreversible and could have severe consequences on the individual's fundamental rights \cite{abomhara2019}.

In order to prevent and mitigate privacy and data protection risks, it is paramount that data controllers comply with the European legal framework designed to protect the privacy and the personal data of individuals.  It is worth mentioning that for data controller(s), it is very important to consider all laws and rules related to the developed technology. However, it is extremely difficult for the data controller(s) to implement a single technological solution or practice that must likely to be compliant with all Member States' laws and rules. Thus, the aim of this paper is to guide data controllers through a general overview of the best practices for privacy and data protection related to personal data processing. During the SMILE research and development, we investigated and critically conducted a theoretical analysis of privacy and data protection by design legal obligation \cite{voigt2017eu,jasmontaite2018data} in the European frameworks and how it can be implemented in the context of biometrics data processing at border control. 
The focus of this paper is only on the European Union (EU) legal obligation established by the General Data Protection Regulation (GDPR) \cite{voigt2017eu}.  

In this paper, we present a brief summary of privacy and data protection best practices to provide more guidance for data controllers and developers on how to comply with the legal obligation for data protection. These  best practices and considerations are based on the lessons learned from the SMILE project.

The remaining part of this paper is organized as following: Section \ref{sec:framework} presents the proposed SMILE data governance framework including a brief summary of EU legal frameworks for data protection, privacy and data protection requirements, organizational and technical measures and data protection impact assessment. Section \ref{sec:best} describes privacy and data protection measures that shall be considered by controllers. Section \ref{sec:conclusions} concludes the paper.

\section{SMILE data governance framework}
\label{sec:framework}

The proposed framework for personal data processing in SMILE consists of multiple elements including the legal frameworks, entities, data protection impact assessment and compliance assessment. 


\subsection{Legal frameworks: EU and national laws}
\label{sec:Legalframework}

The EU has taken numerous specific legislative initiatives with regard to data protection. Currently, the most important instruments is Regulation 2016/679 (GDPR)~\cite{voigt2017eu} on the protection of natural persons with regard to the processing of personal data and on the free movement of such data.  In additional to GDPR, most of other EU legislative initiatives are in the form of directives (e.g., Directive (EU) 2016/680 \cite{sajfert2019data}) which have been implemented or transposed into national laws. This process of implementation allows Member States for some variation along national lines whilst preserving the essential context of the directive. 
Moreover, the European Data Protection Board (EDPB\footnote{The European Data Protection Board is an independent European body whose purpose is to ensure consistent application of the General Data Protection Regulation and to promote cooperation among the EU's data protection authorities.}) gives expert advices regarding data protection in forms of standards and guidelines such as opinions expressed by the Article 29 Data Protection Working Party. This paper focuses only on the legal framework for privacy and personal data protection (GDPR).  Whilst the GDRP is a regulation (which does not require transposition to have legal effect), it still gives a room for Member States to maintain or introduce further conditions, including limitations, with regard to the processing of genetic data, biometric data or data concerning health. 

Biometrics data are considered as "special categories of data" \cite{voigt2017eu}. Therefore, the GDPR prohibits the processing of biometrics data in principle. However,
the Regulation specifies a list of exemptions. In fact, processing biometrics data is covered by certain exemptions listed in Article 9(2) of GDPR. 
Firstly, there is the consent of the data subject, which must be specific, informed, given freely and explicit (Article 9(2)(a) and Article 7 of GDPR). Most of the time, the processing of biometrics data at the borders is subjected to the express consent of the travelers and it is not necessarily written. Moreover, biometrics data may be processed because of the vital interests of the data subject (travelers) or another natural person. Another possible exemption occurs when the processing
is necessary for substantial public interest (Article 9(2)(g) of GDPR). For example, the COVID-19 pandemic forced Member States to reintroduce border checks on Schengen internal borders.  In the end, the GDPR includes the exception for the processing related to the public interest when the data are manifestly made public by the data subject (Article 9(2)(e) of GDPR).

\begin{figure}[t]
\includegraphics[width=\textwidth]{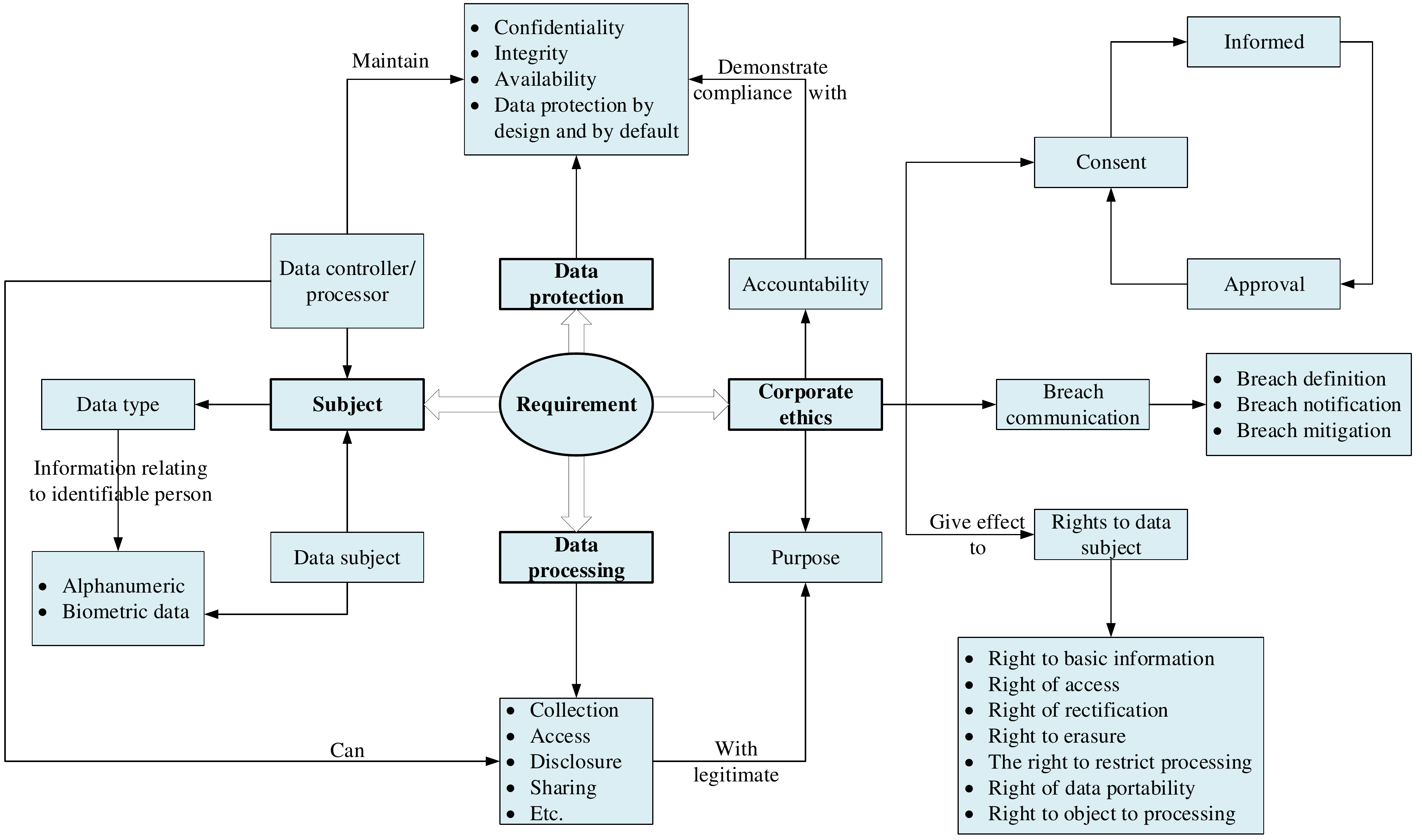}
\caption{Data protection requirements} 
\label{fig2}
\end{figure} 

Based on lessons learned during the SMILE research and development, it is argued that in the context of border crossing, even if the processing is not entirely founded on consent, the travelers have the right to take a role when and how their personal data are used. In Articles 12-23 of GDPR, the law confer on data subject's rights and prescribes the other general rules to be complied with when personal data are processed.  For example, according to Article 25 of GDPR, data controllers processing personal data shall implement Data Protection by Design and by Default (DPbD) measures \cite{voigt2017eu}. Moreover, to ensure compliance with all relevant legal provisions, Data Protection Impact Assessment (DPIA) \cite{bieker2016process,kloza2017data} is required in case of the processing on a large scale of special categories of personal data (Article 35 of GDPR). 

Biometric data processing at border control must comply with privacy and data protection provisions as prescribed in GDPR and Directive (EU) 2016/680 as well as national law. As we mentioned above, the focus of this paper is only GDPR requirements, Figure \ref{fig2} demonstrates set of requirements refer to the processing operations that the data controller must carry out during and after data collection.  Moreover, Table \ref{DPR} presents data protection requirements derived from GDPR. 
 
{\scriptsize
\begin{longtable}{|p{0.8cm}|p{2.2cm}|p{8.8cm}|}
\caption{Data protection requirements derived from GDPR.\label{DPR}}\\
\hline
 \textbf{ID} & \textbf{Requirement} &  \textbf{Biometrics System} \\
\hline
\endfirsthead
\hline
\textbf{ID} & \textbf{Requirement} &  \textbf{Biometrics System}\\
\hline
\endhead 
\endfoot
\hline
\endlastfoot
 Req01 & \textbf{Legal basis} & \tabitem Data processing shall be carried out in accordance with data protection law (e.g., consent, legal obligation of the controller, fair, lawful and transparent processing).  \\

\hline
 Req02 & \textbf{Consent} & \tabitem When applicable, data must collected and processed with freely given, specific, informed and unambiguous consent from the data subject and protection of its vital interests when the data subject is physically or legally incapable of giving consent. \\

\hline
Req03 & \textbf{Purpose limitation} &  \tabitem Biometrics data must be processed for specific, explicitly defined and legitimate purpose and data should not be used for incompatible purposes.  \\

\hline
 Req04 & \textbf{Data minimization} & \tabitem Personal data collected must be adequate, relevant and limited to what is necessary for border control, and in relation to the purpose(s) (Req03) for which those data are processed. \\

\hline
 Req05 & \textbf{Data accuracy} & \tabitem	Biometrics data (and other personal data) must be accurate and, when necessary, kept up to date during the enrollment and matching to avoid false acceptance and/or rejection. Every reasonable step shall be taken to ensure that inaccurate personal data are either erased or rectified with-out delay.\\

\hline
 Req06 & \textbf{Data storage limitation} & \tabitem Unless there is a legal basis (Req01) with appropriate safeguards (i.e., security (Req07) and privacy control) in place, personal data must not be stored more than what is necessary for. \\

\hline
 Req07 & \textbf{Data security} &  \tabitem Data controller(s) implement appropriate technical and organizational measures to ensure a level of security appropriate to the risk guarantee availability, integrity and confidentiality of personal data.  \\

\hline
 Req08& \textbf{Right of information} & \tabitem Data subject (traveler) must have the right to obtain information about their personal data as described in Article 12, 13 and 14 of GDPR \\

\hline
 Req09 & \textbf{Right to access} & \tabitem Data subject (traveler) must have the right to request access to data related to him or her as described in Article 15 of GDPR. Keep in mind, in case of border management,  such a right can be limited or restricted to some extent as stated in Article of 15 of Directive (EU) 2016/680. \\

\hline
 Req10 & \textbf{Right to rectification} & \tabitem  Data subject (traveler) must  have the right to require a controller to rectify any errors in their personal data and o have incomplete personal data completed as described in Article 16 of GDPR\\

\hline
 Req11 & \textbf{Right to erasure} & \tabitem Data subjects must have the right to require a controller(s) to delete their personal data as described in Article 17 of GDPR.  \\

\hline
Req12 & \textbf{The right to restrict processing} & \tabitem Data subjects must have the right to restrict the processing of personal data when accuracy of the personal data is contested, processing is unlawful, data no longer needed, or data subject has objected to processing as described in Article 18 of GDPR.  \newline \tabitem Controllers must inform the data subject in writing of any refusal of rectification (Req10) or erasure of personal data (Req11) or restriction of processing (Req12) and of the reasons for the refusal (Article 19 of GDPR). \\

\hline
 Req13 & \textbf{Right of data portability} & \tabitem Where applicable, data subjects must  have the right to object, on grounds relating to their particular situation, to the processing of personal data as described in Article 20 of GDPR. \\

\hline
 Req14 & \textbf{Right to object the processing} & \tabitem Data subject must have the right to transfer their personal data between controllers (e.g., to move account details from one SMILE platform to another) as described in Article 21 of GDPR. \\

\hline
 Req15 & \textbf{Data disclosure to third parties/countries} & \tabitem In the case where the data needed to disclosure to third parties/countries, data controllers must en-sure (1) an adequate level of data protection before transmission and disclosure, (2) an appropriate safeguards to ensure adequate protection level provided by the third party, (3) an explicit legal permission and (4) the data subject's well informed about the identity of controller(s) in the third parties/countries (Req08).\\

\hline
 Req16 & \textbf{Data breach communication} & \tabitem  Data controller must ensure an assessment of data incidents and prompt notification of breach to data subjects when there is a high risk to the rights and freedoms of natural persons and, with respect to supervisory authorities, notification when the breach is likely to result in a risk to the rights and freedoms of natural persons.\\

\hline
 Req17& \textbf{Accountability}  & \tabitem  Data controller must be able to demonstrate compliance with the data protection principles. \newline \tabitem  Data controller and data processor must take the necessary measures to give effect to the basic principles of data protection set out in GDPR. \newline \tabitem Data controller must be accountable for complying with measures, which give effect to the principles stated above. \newline \tabitem Data controller must carry out "Data Protection Impact Assessment (DPIA)" with accordance to Article 35 of GDPR.
 \\

\end{longtable}
}

\subsection{Entity: Organizational and technical measures}
\label{sec:entity}

The entity is a public or private operator of essential services. In our case, the entity is a competent authority (border/policy authority) composed of social unit of people and complementary partners (another entity) who work together to fulfill and regulate the objectives of the movement of people, animals, and goods across the borders (Regulation (EU) 2016/399 \cite{SBC2016}). All border authority have a management structure that determines relationships between the different activities and the personnel's roles, responsibilities and authority to carry out different tasks. In order to guarantee an effective and efficient interaction and information sharing within and among entities and avoid any management difficulties, this section aims to define responsibilities and allocate roles related to data sharing in accordance with legal requirements (section \ref{sec:Legalframework}) and the entity's policy.

The organizational and administrative measures are expressions of data protection by design approach. Many data protection principles in GDPR are related to organizational measures: fairness, transparency, accuracy, confidentiality and accountability. The data controller should provide evidence that the processing is privacy friendly and in accordance to the legal frameworks.  In the case of border control, it may be important to remember that there may be more than one data controller. Depending on exact circumstances,  the competent authority might be a border authority and/or other law enforcement agencies (LEAs) where they must decide how the data is to be processed etc.  Other entity such as cloud service providers might be classified as data processors. In this case, it is important for the relevant entity (competent authority) to enter into a contract with any other joint controller and processors in order to ensure that the requirements of the GDPR and other national or EU data protection laws (e.g., Directive (EU) 2016/680) are met. Also, the data controller(s) must take all measures needed to ensure that any data processor(s) and other joint controllers are indeed able to fulfil their requirements under such contract. The role of data controllers and data processors are as follows:

\begin{itemize}
	\item Defining data collection purposes, scope, and procedures.
\item	Defining policies for data classification (security levels) and data access control. 
\item Defining the data breach reporting procedures and plans for incident response and disaster recovery. 
\item	Perform the Data Protection Impact Assessment (DPIA) -- discussed in Section \ref{sec:DPIA}. 
\item 	Design, create, and implement IT processes and systems that would enable the data controllers to gather personal data.
\item 	Define the used tools and strategies to gather personal data.
\item 	Implement security measures that would safeguard personal data.
\item	Transfer data from the data controller to another entity (can be data controller or data processor) and vice versa.

\end{itemize}

In the case of joint controllers or data is begin processed by a data processor(s), Article 28 of the GDPR lays down requirements that must be in place between a controller(s) and processor(s), in order to protect the rights of the data subject. Data controller shall ensure that processors are authorized to process the personal data and have committed themselves to confidentiality or are under an appropriate legal obligation of confidentiality (Articles 27-29 of the GDPR). Joint data controllers agreement must be in place and signed by data controllers and processors. The main scope of the joint data controllers and processors agreement are (in accordance to Article. 26 of the GDPR):  

\begin{itemize}
	\item The agreement must lay down the distribution of responsibilities among the joint controllers and processors in connection with all the processing of  personal data.
\item	The agreement must lay down the rules of sharing and transferring of the personal data.
\end{itemize}

Moreover, controllers must ensure the data subject's rights are being upheld within the entity/system.  Data controllers and processors are required to respect data subject' rights as defined in the GDPR (Articles 12-23 of GDPR). The data controller should give clear and documented instructions to data subjects about how to excise their rights (Req08 to Req13) as presented in Table \ref{DPR}. Also, the data controller may ask an opinion of a privacy expert to ensure compliance with the law. A data protection officer (DPO) shall be designated where the core activities of the data controller consists of processing on a large scale of special categories of personal data such as biometrics data.  The DPO, who may or may not be the same person for several data controllers should (among others) provide advice and guidance to the controllers and processors on the requirements of the data protection, provide advice about the DPIA as well as monitor and support its performance and be the person of contact for data subjects and for consulting with national supervisory authorities. 

Table \ref{table2} presents a summary of the organizational and technical measures for biometrics data processing.  The measures presented in Table \ref{table2} are considered based on ISO/IEC 27001:2013 standard \cite{international2013iso}. ISO/IEC 27001:2013 specifies the requirements for establishing, implementing, maintaining and continually improving an information security management system within the context of the organization. 

{\scriptsize
\begin{longtable}{|p{1.8cm}|p{8cm}|p{2cm}|}
\caption{Organizational and technical measures.\label{table2}}\\
\hline
\textbf{Name of measure} & \textbf{Description} &  \textbf{Reference to ISO/IEC 27001:2013} \\
\hline
\endfirsthead
\hline
\textbf{Name of measure} & \textbf{Description} &  \textbf{Reference to ISO/IEC 27001:2013}\\
\hline
\endhead 
\endfoot
\hline
\endlastfoot
 Policy and procedures & \tabitem Data controllers must document its security and privacy policy for privacy and data protection. \newline \tabitem The  policy must be reviewed and revised by the all data controllers (in the case of joint controllers).  & \tabitem A.5 Security policy.  \\

\hline 
Roles and responsibilities & \tabitem Data controllers must  specify and allocate role(s) and responsibilities related to data processing  in accordance with the policy and procedures.   \newline \tabitem Data controllers must (when applicable) appoint a Data Protection Officer (DPO) and define the role of the authorized persons to reduce opportunities for unauthorized or unintentional modification or misuse of personal data.  & \tabitem A.6.1.1 Information security roles and responsibilities
 \newline \tabitem A6.1.2 Segregation of duty \\

\hline 
Access control policy & \tabitem Access control policy must define the rights of access to each role(s) in the competent authority. \newline \tabitem  An appropriate access control mechanism (e.g., Role based access control \cite{ferraiolo2001proposed}) must be implemented to restrict access rights for specific user roles based on need-to-know principle \cite{janczewski2000need}. & \tabitem A.9.1 Business requirement of Access control policy \newline \tabitem A.9.2 User access management  \\ 

\hline 
Resource \& asset management & \tabitem System hardware, software and network resources must be reviewed and approved by controllers before any resource is put in action. \newline \tabitem Resources must be classified by the their sensitivity to limit unauthorized disclosure/modification of any sensitive information/data.   & \tabitem  A.8 Asset management
\newline \tabitem A.8.2.1 Classification of information 
\\

\hline 
Data \& controllers or processors & \tabitem  An agreement must define the role and responsibilities of each data controller and processor with respect to confidentiality, non-disclosure etc. & \tabitem  A.15.1.1 Information security policy for supplier relationships \newline \tabitem A.15.1.2 Addressing security within supplier agreements.  
\\ 

\hline 
Incidents handling \& personal data breaches & \tabitem Incident response plan with detailed notification procedures for reporting must be defined to ensure effective and orderly response to incidents pertaining personal data. & \tabitem A.16 Information security incident management \\

\hline 
Human resources security & \tabitem  Entities must ensures that responsibilities and obligations related to the processing of personal data are clearly communicated to its personnel.  \newline \tabitem Entities must ensure that its personnel involved in the date processing are well trained and understand the  policy related to confidentiality and non-disclosure. 
& \tabitem  A.7 Human resource security \newline \tabitem A.7.2.2 Information security awareness, education and training
\\

\hline 
Security risk assessment \& DPIA & \tabitem  Data controllers must ensure the performance of security risk assessment and the performance of DPIA to map data protection and privacy requirements (described in Table \ref{DPR}) to threats, vulnerabilities, risks and mitigation measures for development.   & \tabitem   A12.6 Technical vulnerability management \newline \tabitem A.14.2 Security in development and support processes
\\

\hline 
Activity \& event logging and monitoring  & \tabitem Data controllers must ensure logging and auditing record  of authorized users' activities and events (read, write, view etc.) with timestamped and adequately protected against tampering and unauthorized access.  \newline \tabitem Action (collection, deletion, disclosure etc.) and system operators and system administrators must be logged timestamped and adequately protected against tampering and unauthorized access. & \newline \tabitem A.12.4 Logging and monitoring \\

\hline  
Data security  & \tabitem  Data controllers must ensure protecting digital data from destructive forces and from the unwanted actions of unauthorized users, such as a cyberattack or a data breach. & \tabitem A.10.1 Cryptographic control \newline \tabitem A.12 Operations security \\

\hline 
Backup & \tabitem Data controllers keep back-up copies in a locked and fire-proof facility and kept separate from operating equipment. \newline \tabitem Back-ups copy shall be protected against malware and incidents. & \tabitem A.12.3 Back-Up \\

\hline 
Data deletion \& disposal & \tabitem Personal data must be deleted when no longer needed. This includes shredding of paper and portable media used to store personal data.  & \tabitem A. 8.3.2 Disposal of media \newline \tabitem A.11.2.7 Secure disposal or re-use of equipment
\\

\end{longtable}
}

\subsection{Data protection impact assessment (DPIA)}
\label{sec:DPIA}

Article 35 of the GDPR introduces the necessity of DPIA.  It is a process that helps to identify and minimize the privacy and data protection risks resulting from the processing of personal data [1, 2].  The process is designed to describe the processing, assess its necessity and proportionality and help manage the risks to the rights and freedoms of natural persons.  Article 25 GDPR establishes that, both at the time of the determination of
the means for processing and at the time of the processing itself, the controller shall implement appropriate technical and organizational measures which are
designed to implement data protection principles in an effective manner and to integrate the necessary safeguards into the processing. The DPIA
may be considered as an organizational measure. Thus, the DPIA helps data controllers to comply with legal requirements of data protection and demonstrate the appropriate measures where it is used to check compliance against data protection regulation.  To ensure compliance with legal requirements of data protection and demonstrate the appropriate measures, DPIA shall include: 

\begin{itemize}
	\item A systematic description of the processing activities.
 \item A description of the purpose for the processing of personal data.
 \item An assessment of whether or not the processing of personal data is necessary and proportionate to the purpose. 
 \item An assessment of the privacy and data protection risks for the data subject. 
 \item Planned privacy and data protection risks mitigation in order to safeguard data and protect privacy.
\end{itemize} 
 
Controllers shall consult with the DPO, if such an officer has been designated, in connection with the performance of DPIA. In case the processing of biometrics data and personal data entail a high risk which cannot be mitigated through reasonable measures, controllers shall request an advance discussion with the Data Protection Authority (DPA) before the processing is commenced.

It is argued that the DPIA is a preliminary step of any privacy and data protection by design  process \cite{cavoukian2009privacy,jasmontaite2018data}. The loss of confidentiality, integrity and availability of data concerning biometrics data processing is a high risk. Once the risks have been identified, the appropriate solutions solutions developed according to PbD principles should balance and take into account state-of-the-art of the technology and the costs of implementation. The controller shall take into account the risks of varying likelihood and severity for rights and freedoms of natural persons posed by the processing. However, the management of the data processing and the risk assessment are crucial. 
The report "Privacy and Data Protection by Design, from policy to engineering" sets out some strategies for the implementation and defines eight PbD strategies and three data protection goals \cite{danezis2015privacy}. These recommendations are strictly related to the Hoepman et alia's PbD strategies \cite{colesky2016critical}.  Table 2 provides an overview of the privacy by design (PbD) strategies for biometrics data processing and the possible implementation measures in each of the phases of the data processing \cite{danezis2015privacy,d2015privacy}.  A brief overview of the strategies is as follows: 

\begin{itemize}
	\item \textbf{Inform}: Data subject should be adequately informed whenever his/her data is processed (transparency).
	\item \textbf{Control}: Data subjects should be provided control over the processing of their personal data (rights to data subject).
	\item \textbf{Minimize}: The amount of personal data should be restricted to the minimal amount possible (data minimization). 
	\item \textbf{Hide}: Personal data and their interrelations should be hidden,  not communicated in  plain text. 
	\item \textbf{Separate}: Personal data should be processed in a distributed fashion, in separate compartments whenever possible. Personal data should be stored in separate databases and areas for each purpose and process.  
	\item \textbf{Aggregate}: Personal data should be processed at the highest level of aggregation and with the least possible detail in which it is (still) useful. This would ensure the enforceability of the data subject’s rights, without prejudice to the business value and purpose of the collection and use.
	\item \textbf{Enforce}: A privacy policy compatible with legal requirements (e.g., GDPR requirements) should be in place and should be enforced.
	\item \textbf{Demonstrate}: Data controllers must be able to demonstrate compliance with privacy policy into force and any applicable legal requirements. 
\end{itemize}

{\scriptsize
\begin{longtable}{|p{2.7cm}|m{2cm}|p{7.3cm}|}
\caption{Security and data protection measures.\label{table2}}\\
\hline
\textbf{Processing phase} & \textbf{PbD strategies} &  \textbf{Implementation} \\
\hline
\endfirsthead
\hline
\textbf{Processing phase} & \textbf{PbD strategies} &  \textbf{Implementation}\\
\hline
\endhead 
\endfoot
\hline
\endlastfoot
    \multirow{5}{*}{} & Inform & \tabitem Controller(s)  must provide appropriate information to data subject about the data collection and the purpose of the data. \newline \tabitem Controller(s) must use transparency mechanisms (whenever possible) to inform data subject about the processing. \newline \tabitem Controller(s) must provide contact point that data subject can use to practice data subject rights. \newline \tabitem Controller(s) must use multiple languages if necessary and enrich information to use photographs, audio, video, etc (when applicable). 
  \\ \cline{2-3}
    & Minimize & \tabitem Controller(s) must define what data are needed/necessary before collection to reduce data fields, define relevant controls and avoid collection of unwanted information.  \\ \cline{2-3}
		
		 & Hide & \tabitem  Controller(s) must implement privacy enhancing techniques, e.g. anti-tracking techniques, encryption techniques, identity masking techniques, secure file sharing techniques, etc. to avoid unnecessary exposure of data. \\ \cline{2-3}
		
		Data collection & Aggregate & \tabitem Controller(s) should use anonymization/pseudonymization whenever possible. \newline \tabitem Controller(s) should remove unnecessary and excessive information. \\ \cline{2-3}
		
		 &  Control & \tabitem  Controller(s) must implement an appropriate mechanism for data subject to express their rights includes informed consent, rights to withdraw consent, rights to give access for the rectification, blocking, or deletion of personal data and rights to submit questions or complaints relating to data protection and security. \newline \tabitem Controller(s) must have/implement mechanisms for expressing privacy preferences (e.g., which biometric data an individual prefer to use).
 \\ \cline{2-3}
		
		 &  Demonstrate & \tabitem Controller(s) must demonstrate that they have defined what data to be collected, why and how including documentation demonstrating the system design and security (auditing re-ports, vulnerability scanning, data breach management, etc.). \\
    \hline 
		
		\multirow{3}{*}{} & Hide & \tabitem Controller(s) must encrypt data at rest or in transit. \newline \tabitem Controller(s) must use authentication and access control mechanisms to process (e.g., access, read, write, copy etc.) data.  \newline \tabitem Controller(s) must use other measures (e.g., encrypted backups) for secure data storage. 
 \\ \cline{2-3}
   Data storage & Separate & \tabitem Controller(s) must use distributed/ decentralized storage and analytic facilities whenever possible. \newline \tabitem Controller(s) separate sensitive personal data from less sensitive personal data (in the database, access to sites, for clients and units, etc).
  \\ \cline{2-3} 
	
	& Enforce/ Demonstrate &  Controller(s) must demonstrate what type of security and privacy techniques are used as well as how these techniques are enforced?\\
	\hline
	
	\multirow{2}{*}{Data use \&  disclosure} & Aggregate & \tabitem Controller(s) must use anonymization/pseudonymization techniques whenever possible. 
 \\ \cline{2-3}

& Hide & \tabitem  Controller(s) must use searchable encryption and privacy-preserving computations, whenever possible.\\	
\end{longtable}
}


\section{Privacy and data protection best practices}
\label{sec:best}
This section describes privacy and data protection best practices that shall be considered by controllers. Privacy and data protection measures shall be chosen on the basis of risk assessments and DPIA.  In order to guarantee privacy and data protection, this paper presents the aspects of data protection that relate most directly to biometric data processing in border management. As discussed above, there are many factors to consider when it comes to data protection such as: 

\begin{enumerate}
	\item \textbf{Compliance with security requirements:} Privacy and data protection requirements deal with security issues, such as confidentiality, integrity and availability when processing personal data.  Activities required to preserve confidentiality, integrity  and availability including granting access only to authorized personnel, applying encryption to data that will be sent over the Internet or stored on digital media, performing risk assessments to uncover new vulnerabilities, building software defensively and developing a disaster recovery plan to ensure that the business can continue to exist in the event of a disaster or loss of access by personnel. Controllers ought to take the following into consideration: 
	\begin{itemize}
	\item \textbf{Confidentiality:} Controllers must ensure the security of collected data and be able to prevent data leakage. Moreover, controllers must have an access control capability that can authenticate users who want to access data and authorize eligible users to have access.
\item \textbf{Integrity:} Controllers must be able to prevent data loss and any unauthorized modifications of data as well as to verify the integrity and authenticity of the collected data. 
 \item \textbf{Availability:} Controllers must ensure data backup to prevent loss of data due to natural disasters (fire, flooding, storms, earthquakes, etc.) or human actions such as Denial of Service (DoS) attacks. 

 \item \textbf{Auditing:} Controllers must allow security and data protection audits as a systematic evaluation of the security of a system hardware and software by measuring how well it conforms to a set of established criteria. 
\end{itemize}

\item \textbf{Compliance with the regulation requirements}:  As explained above, for border control related purposes data protection compliance essentially refers to the GDPR and Directive (EU) 2016/680. Controllers must take into consideration all the requirements and main principles presented in the GDPR and Directive (EU) 2016/680. These include:

\begin{itemize}
	\item \textbf{Collection limitation principle:}  Aimed at limiting the collection of personal data. 
	
\item \textbf{Lawfulness of the data processing principle:} Require for data to be obtained by lawful and fair means and when appropriate, with the individual's knowledge or consent.

\item \textbf{Accountability principle:} Requires data controllers to adhere to applicable legislation themselves, at their own initiative and best efforts, and to be able to demonstrate such compliance whenever needed.

\item \textbf{Transparency principle: } Aimed at strengthening data subjects' position while defending their right to personal data protection.

\item \textbf{Right to data subject principle:} Data subjects can make a specific request and be assured that his/her personal data is not being misused for other than the legitimate purpose for which it was originally provided. GDPR empowers data subjects with rights including the right to information, right to access, right to rectification and erasure as well as the right to be forgotten, to name a few (Article 12-23, GDPR).

\end{itemize}

\end{enumerate}

\section{Conclusions and recommendations}
\label{sec:conclusions}

Data protection aims at guaranteeing the individual's right to privacy. It refers to the technical and legal framework designed to ensure that personal data are safe from unforeseen, unintended or unauthorized use. Data protection therefore includes e.g., measures concerning collection, access to data, communication and conservation of data. In addition, a data protection strategy can also include measures to assure the accuracy of the data. In the context of biometrics data processing, data protection issues arise whenever data relating to persons are collected and stored.

Privacy and data protection by design  are currently mandatory with the GDPR, but many entities still find difficulties with the concept, both in terms of what it exactly means and how to implement it as a system quality attribute. Moreover, the law imposes high administrative fines in case of infringements (Article 83 of GDPR). For these reasons, in the future, the aim  is to investigate more about privacy and data protection measures and propose a comprehensive privacy and data protection management framework for biometrics data processing. Future work could also focus on analyzing if and how far our proposed  privacy and data protection management framework could comply and adhere to other legislation.

\section*{Acknowledgements} 
This work is carried out in the EU-funded project SMILE (Project ID: 740931), [H2020-DS-2016-2017] SEC-14-BES-2016 towards reducing the cost of technologies in land border security applications.

%
%
%
\bibliographystyle{splncs04}
 \bibliography{Bibliography}

\end{document}